\documentclass{aastex6}

\usepackage{graphicx}
\usepackage{amsmath}
\usepackage{tikz}

\begin{document}

\title{Lorentz Invariance Violation Effects on Gamma-Gamma Absorption and Compton Scattering}
\author{Hassan Abdalla\altaffilmark{1,2} \footnote{hassanahh@gmail.com} and Markus B\"{o}ttcher\altaffilmark{1} \footnote{Markus.Bottcher@nwu.ac.za }} 

\altaffiltext{1}{Centre for Space Research, North-West University, Potchefstroom 2520, South Africa}
\altaffiltext{2}{Department of Astronomy and Meteorology, Omdurman Islamic University, Omdurman 382, Sudan}

\begin{abstract}
In this paper, we consider the impact of Lorentz Invariance Violation (LIV) on the $\gamma-\gamma$ opacity of the Universe
to VHE-gamma rays, compared to the effect of local under-densities (voids) of the Extragalactic Background Light, and on the 
Compton scattering process. Both subluminal and superluminal modifications of the photon dispersion relation are considered.
In the subluminal case, LIV effects may result in a significant reduction of the $\gamma-\gamma$ opacity for 
photons with energies $\gtrsim 10$~TeV. However, the effect is not expected to be sufficient to explain the 
apparent spectral hardening of several observed VHE $\gamma$-ray sources in the energy range from 100 GeV -- a few TeV,
even when including effects of plausible inhomogeneities in the cosmic structure. Superluminal modifications of the 
photon dispersion relation lead to a further enhancement of the EBL $\gamma\gamma$ opacity. We consider, for the first 
time, the influence of LIV on the Compton scattering process. We find that this effect becomes relevant only for photons 
at ultra-high energies, $E \gtrsim 1$~PeV. In the case of a superluminal modification of the photon dispersion relation,
both the kinematic recoil effect and the Klein-Nishina suppression of the cross section are reduced. However, we argue 
that the effect is unlikely to be of astrophysical significance. 
\end{abstract}


\section{Introduction} \label{sec:intro}

Recent astronomical observations and laboratory experiments appear to show hints that several phenomena in physics, 
astrophysics and cosmology oppose a traditional view of standard-model physics \citep[e.g.,][]{Riess98, Furniss13}. 
This has motivated developments of modified or alternative theories of quantum physics and gravitation 
\citep[e.g.,][]{Capozziello13,W14,Nashed14,Arbab15,Sami18}, generally termed {\it physics beyond the standard 
model} \citep[e.g.,][]{Sushkov11,Abdallah13,El-Zant15}.

The special theory of relativity postulates that physical phenomena are identical in all inertial frames. Lorentz
invariance is one of the pillars of modern physics and is considered to be a fundamental symmetry in Quantum Field 
Theory. However, several quantum-gravity theories postulate that familiar concepts such as Lorentz invariance may 
be broken at energies approaching the Planck energy scale, $E_P \sim 1.2 \times 10^{19} GeV$ 
\citep[e.g.,][]{Amelino, Jacob08, Liberati09,Amelino13,Tavecchio16}. Currently such extreme  energies are unreachable by 
experiments on Earth, but for photons traveling over cosmological distances the accumulated deviations from Lorentz 
invariance may be measurable using Imaging Atmospheric Cherenkov Telescope facilities, in particular the future
Cherenkov Telescope Array (CTA) \citep[e.g.,][]{Fairbairn14,Lorentz17}. 

A deviation from Lorentz invariance can be described by a modification of the dispersion relation of photons 
and elementary particles such as electrons \citep[e.g.,][]{Amelino,Tavecchio16}. It is well known that the 
speed of light in a refractive medium depends on its wavelength, with shorter wavelength (high momentum) modes 
traveling more slowly than long wavelength (low momentum) photons. This effect is due to the sensitivity of 
light waves to the microscopic structure of the refractive medium. Similarly, in quantum gravity theories, 
very high energy photons could be sensitive to the microscopic structure of spacetime, leading to a violation
of strict Lorentz symmetry. In that case, $\gamma$-rays with higher energy are expected to propagate more 
slowly than their lower-energy counterparts \citep[e.g.,][]{Amelino, Fairbairn14, Tavecchio16,Lorentz17}.   
This would lead to an energy-dependent refractive index for light in vacuum. Therefore, the deviation from Lorentz 
symmetry can be measured by comparing the arrival time of photons at different energies originating from the same 
astrophysical source \citep[e.g.,][]{Amelino,Azzam09, Tavecchio16, Wei17,Lorentz17}.

Gamma rays from objects located at a cosmological distance with energies greater than the threshold energy for 
electron-positron pair production can be annihilated due to $\gamma-\gamma$ absorption by low-energy extragalactic 
background photons \citep{Nishikov62}. The intergalactic $\gamma-\gamma$ absorption signatures have attracted great 
interest in astrophysics and cosmology due to their potential to indirectly measure the Extragalactic Background
Light (EBL) and thereby probe the cosmic star-formation history \citep[e.g.,][] {BW15}. The predicted $\gamma-\gamma$ 
absorption imprints have been studied employing a variety of theoretical and empirical methods 
\citep[e.g.,][]{Stecker69, Stecker92,HD01, Primack2005, Aharonian06,Franceschini08,Razzaque09,Finke10,Dominguez11a,Gilmore12}. 

Recent observations indicate that the very-high-energy (VHE; $E > 100$~GeV) spectra of some distant 
($z \gtrsim 0.5$) blazars, after correction for $\gamma-\gamma$ absorption by the EBL, appear harder 
than physically plausible \citep[e.g.,][]{Furniss13}, although systematic studies of the residuals of
spectral fits with standard EBL absorption on large samples of VHE blazars \citep[e.g.,][]{BW15, Mazin2017}
reveal no significant, systematic anomalies on the entire samples. Nevertheless, the unexpected VHE-$\gamma$-ray 
signatures seen in a few individual blazars are currently the subject of intensive research. 
Possible explanations of this spectral hardening include the hypothesis 
that the EBL density could be lower than expected from current EBL models \citep{Furniss13}, an additional  
$\gamma$-ray emission component due to interactions along the line of sight of extragalactic Ultra-high-Energy 
Cosmic Rays (UHECRs) originating from the blazar \citep[e.g.,][]{EK10,Dzhatdoev15}, the existence of exotic 
Axion-Like Particles (ALPs) into which VHE $\gamma$-rays can oscillate in the presence of a magnetic field, 
thus enabling VHE-photons to avoid $\gamma-\gamma$ absorption \citep[e.g.,][]{Dominguez11b, Dzhatdoev17}, 
EBL inhomogeneities \citep[e.g.,][]{Furniss15,KF16,Abdalla17} and the impact of LIV, which can lead to an 
increase of the $\gamma\gamma$ interaction threshold and thus to a reduction of cosmic opacity (especially 
at energies beyond $\sim 10$~TeV), thus allowing high-energy photons to avoid $\gamma-\gamma$ absorption 
\citep[e.g.,][]{Tavecchio16,Abdalla18}.  

In this paper, we discuss the reduction of the EBL $\gamma-\gamma$ opacity due to the existence of underdense 
regions along the line of sight to VHE gamma-ray sources (including contributions of both the direct star light
and re-processed emission to the EBL) and compare the results with the LIV effect on cosmological photon propagation. 
We consider the LIV effect only for photons, but not for electrons, since the high-energy synchrotron spectrum of 
the Crab Nebula imposes a stringent constraint on any deviation of the electron dispersion relation from Lorentz 
invariance \citep[e.g.,][]{Jacobson03}. 

LIV may also effect the process of Compton scattering, which is likely to be an important $\gamma$-ray production 
process in many astrophysical high-energy sources, such as accreting black hole binaries, pulsar wind nebulae, the 
jets from active galactic nuclei, and supernova remnants. 
In this paper, we discuss, to our knowledge for the first time, the impact of LIV on the Compton scattering process,
both on energy-momentum conservation and on the Compton cross section. 

In Section \ref{void} we investigate the impact of the existence of cosmic voids along the line of sight to a 
distant VHE $\gamma$-ray source, by using the EBL model devoloped by \cite{Finke10}.
In Section \ref{livopa} we review the impact of LIV on the EBL $\gamma\gamma$ opacity. In Section \ref{livcom} 
we investigate LIV effects on the Compton scattering process, starting with basic conservation of energy and momentum, 
using the LIV-deformed dispersion relation for photons. The results are presented in Section \ref{Results}, where we 
compare our results with predictions from standard quantum electrodynamics (QED). We summarize and discuss our results 
in Section \ref{Summary}. Throughout this paper the following cosmological parameters are assumed: 
$H_{0} = 70$~km~s$^{-1}$~Mpc$^{-1}$, $\Omega_{m} = 0.3$, $\Omega_{\Lambda} = 0.7$. 

\section{The impact of a Cosmic Void on the EBL energy density distribution}\label{void}

A generic study of the effects of cosmic voids along the line of sight to a distant astronomical object (e.g. blazar) 
on the EBL $\gamma-\gamma$ opacity has been done in \citep{Abdalla17}. In that paper, the EBL was represented using the 
prescription of \cite{Razzaque09}, taking into account only the direct starlight contribution to the EBL. Assuming 
that a spherical cosmic void with raduis $R$ is located with its center at redshift $z_v$, between an observer and a 
VHE $\gamma$-ray source located at redshift $z_{s}$, the angle- and photon-energy-dependent EBL energy density at each 
point between the observer at redshift zero and the source was calculated. The cosmic void was represented by setting 
the star formation rate to 0 within the volume of the void. We found that the EBL deficit is proportional to the size 
of the void. Therefore, the effect of a number $n$ of voids of radius $R_1$ is aproximately the same as the effect of 
a large void with radius $R_n = n R_1 $.

Since in the \cite{Razzaque09} prescription, only the direct starlight contribution to the EBL is considered, the 
work of \citep{Abdalla17} under-predicts the EBL $\gamma-\gamma$ opacity for VHE $\gamma$-rays with energies of
$E \gtrsim 1$~TeV. Such VHE photons preferentially interact with longer-wavelength (IR -- FIR) EBL photons,
which are dominated by dust re-processing of starlight, which is neglected in \cite{Razzaque09}. To study 
the impact of a cosmic void on the full EBL spectrum, from far infrared through visible and extending into 
the ultraviolet, we used the EBL model of \cite{Finke10}, in which stars that evolved off the main sequence 
and re-emission of absorbed starlight by dust are considered. In all other aspects, we follow the formalism
of \cite{Abdalla17}. 

One of the most complete public catalogues of cosmic voids \cite{Sutter12} is based on data from the Sloan
Digital Sky Survey (SDSS), with effective radii of voids spanning the range $5- 135 h^{-1}$~Mpc. Also, there is 
evidence for a $300 h^{-1}$~Mpc under-dense region in the local galaxy distribution \citep[e.g.,][]{Keenan13}. Recent
measurements of optical and NIR anisotropies \citep[e.g.,][]{Matsuura17}, at $1.1$ and $1.6~\mu m$, indicate
that the resulting amplitude of relative EBL fluctuations is typically in the range of 10 to 30\% \cite{Zemcov14}. 

\begin{figure}[ht]
\plottwo{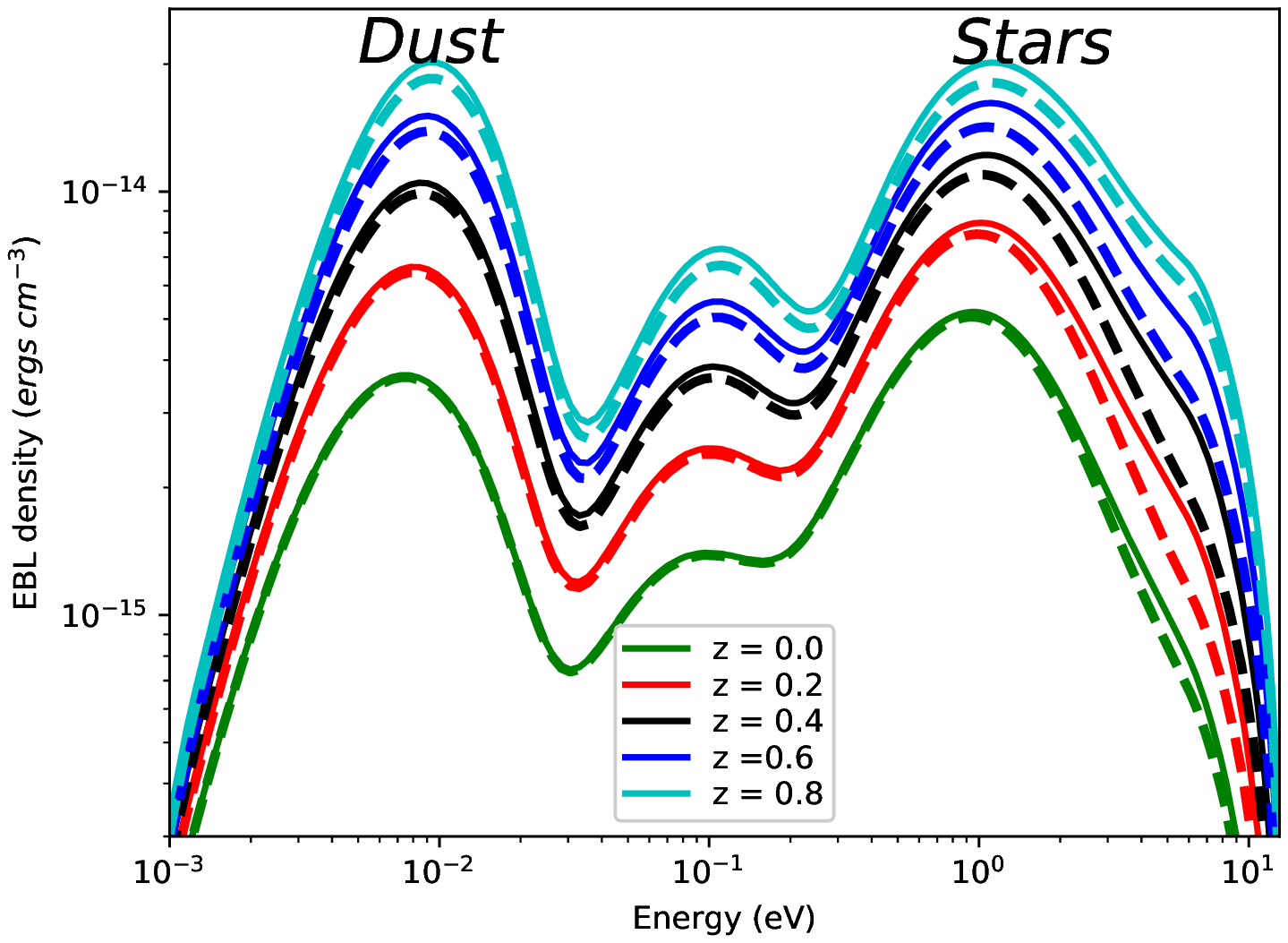}{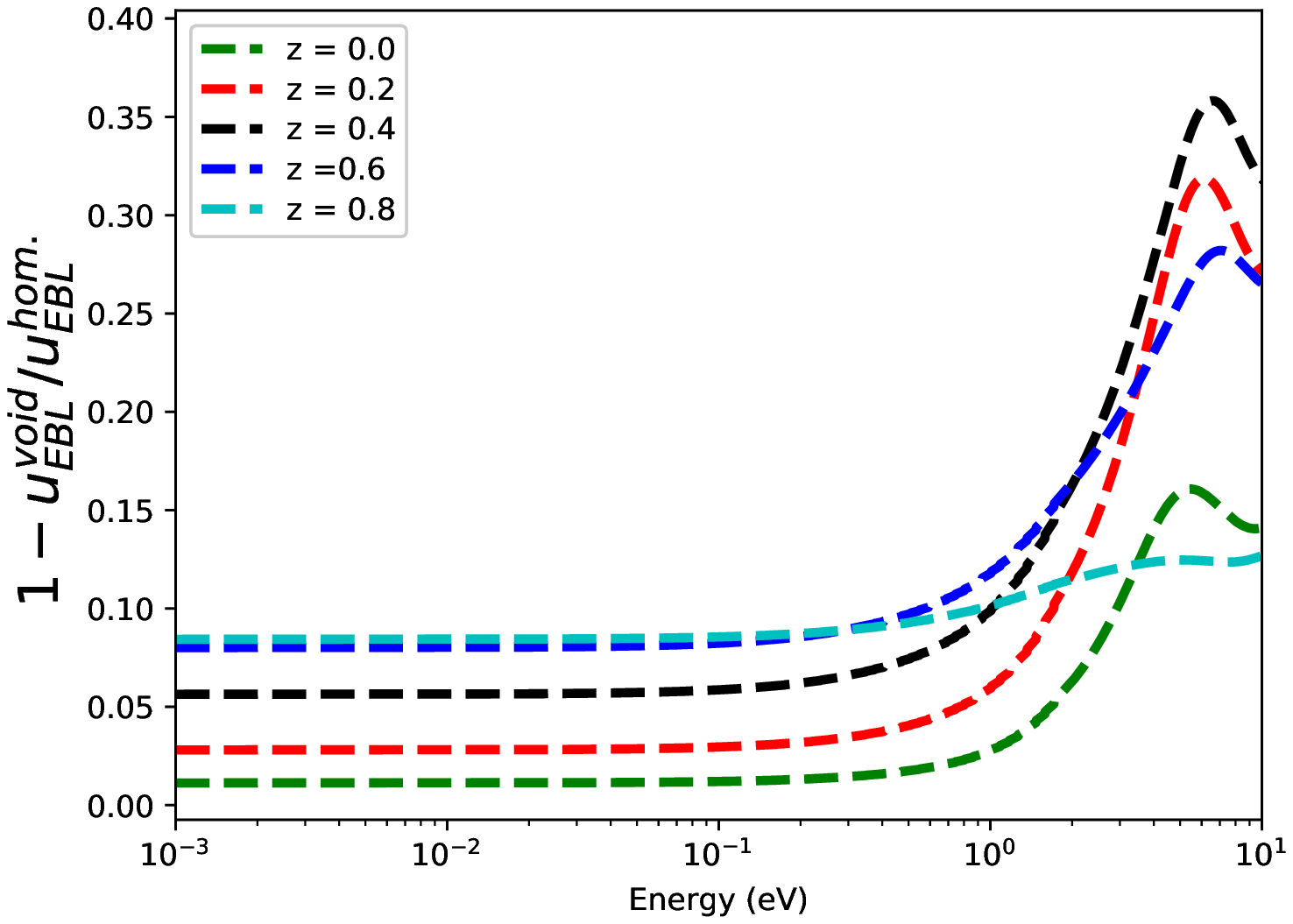}
\caption{Left panel: Differential EBL photon energy density as a function of distance (redshift) along the line of
sight. The solid lines represent the homogeneous case ($R = 0$), and the dashed lines represent the EBL energy density 
considering an accumulation of about 10 voids of typical sizes with radius $R = 100 \, h^{-1}$~Mpc centered at redshift 
$z_v= 0.3$. The EBL energy density increases with redshift because of the star formation 
rate increasing with redshift at low redshifts \citep[e.g.,][]{Cole2001}\\
Right panel: Relativie EBL energy density deficit due to the presence of a void for the same cases
as represented in the left panel. $U^{void}_{EBL}$ and $U^{hom}_{EBL}$ are the EBL energy densities considering 
the cosmic void case and the homogeneous case, respectively. As expected, the maximum EBL energy density deficit occurs around 
the center of the voids ($z=0.3$), and the EBL energy density deficit beyond the center of the cosmic voids is greater than the 
deficit in front of it, comparing points at the same distance from the voids center, due to the star formation rate 
increasing with redshift. 
\label{fig:EBL}
}
\end{figure}

The impact of an accumulation of cosmic voids amounting to a total size of radius $R = 1 \, h^{-1}$~Gpc (where $h
= H_0 / (100$~km~s$^{-1}$~Mpc$^{-1})$) centered at redshift $z_v= 0.3$, is illustrated in Figure \ref{fig:EBL}. 
The EBL energy density spectrum in the presence of voids (dashed lines) is compared to the homogeneous case 
(solid lines) at different points (redshifts, as indicated by the labels) along the line of sight in the left 
panel of Figure \ref{fig:EBL}. The fractional difference between the homogeneous and the inhomogeneous case 
as a function of photon energy for various redshifts along the line of sight is presented in the right panel 
of Figure \ref{fig:EBL}. We notice that the EBL deficit is smaller for low-energy (IR) photons than for 
optical -- ultraviolet photons. This is because the UV EBL is dominated by hot, young stars, thus more 
strongly reflecting the local effect of the void. Since in this work we set only the star formation rate 
inside the void equal to zero, dust re-processing of star light produced outside the void, still takes 
place inside the void. 
As can be seen from Figure \ref{fig:EBL}, with our choice of a void configuration,
the impact of underdense regions is comparable to the measured optical and NIR anisotropy 
\cite{Zemcov14}.
The impact of the EBL deficit due to the cosmic voids on the EBL $\gamma-\gamma$ 
opacity will be presented in Section \ref{EBLresults}.
\section{Lorentz Invariance Violation: Cosmic opacity}\label{livopa}
In this Section we review the imprints of LIV on the cosmic $\gamma-\gamma$ opacity, primarily based on the work 
by \cite{Tavecchio16}. The results will be compared to the imprints of EBL inhomogeneities discussed in the previous
section. The deviation from Lorentz symmetry can be described by a modification of the dispersion relation 
of photons and electrons \citep[e.g.,][]{Amelino,Tavecchio16}:
\begin{equation}\label{dis}
 E^{2} = p^2 c^2 + m^2 c^4  + ~ S ~  E^2 \left(\frac{E}{ E_{LIV}} \right)^{n},
\end{equation}
where $c$ is the conventional speed of light in vacuum, ``$S = -1$'' represents a subluminal scenario 
(decreasing photon speed with increasing energy), and ``$S = +1$'' represents the superluminal case 
(increasing photon speed with increasing energy). The characteristic energy $E_{\rm LIV}$ is parameterized
as a fraction of the Planck energy, $E_{\rm LIV} = E_{P} / \xi_n$, where the dimensionless parameter $\xi_n$ 
and the order of the leading correction $n$ depend on particle type and theoretical framework
\citep[e.g.,][]{Amelino,Tavecchio16}. A value of $E_{LIV} \sim E_P$ (i.e., $\xi_1 = 1$)
has been considered to be the physically best motivated choice \citep[e.g.,][]{Liberati09, Fairbairn14, Tavecchio16} 
This is consistent with the results of \cite{BW15} which constrained  $E_{LIV} >  0.65 ~ E_P$. Some
authors \citep[e.g.,][]{Schaefer98, Biller99} argue that the best constraint from current data is $\xi_{1} \leq O(1000)$. 

In the literature \cite[e.g.,][]{Tavecchio16}, usually only the 
subluminal case is considered for the LIV effect on $\gamma\gamma$ absorption, as this is the case that
could lead to an increase of the $\gamma\gamma$ interaction threshold and consequently, a decrease of the 
opacity. In this work, for completeness, we consider both the subluminal and superluminal cases. 

\begin{figure}[ht]
\plottwo{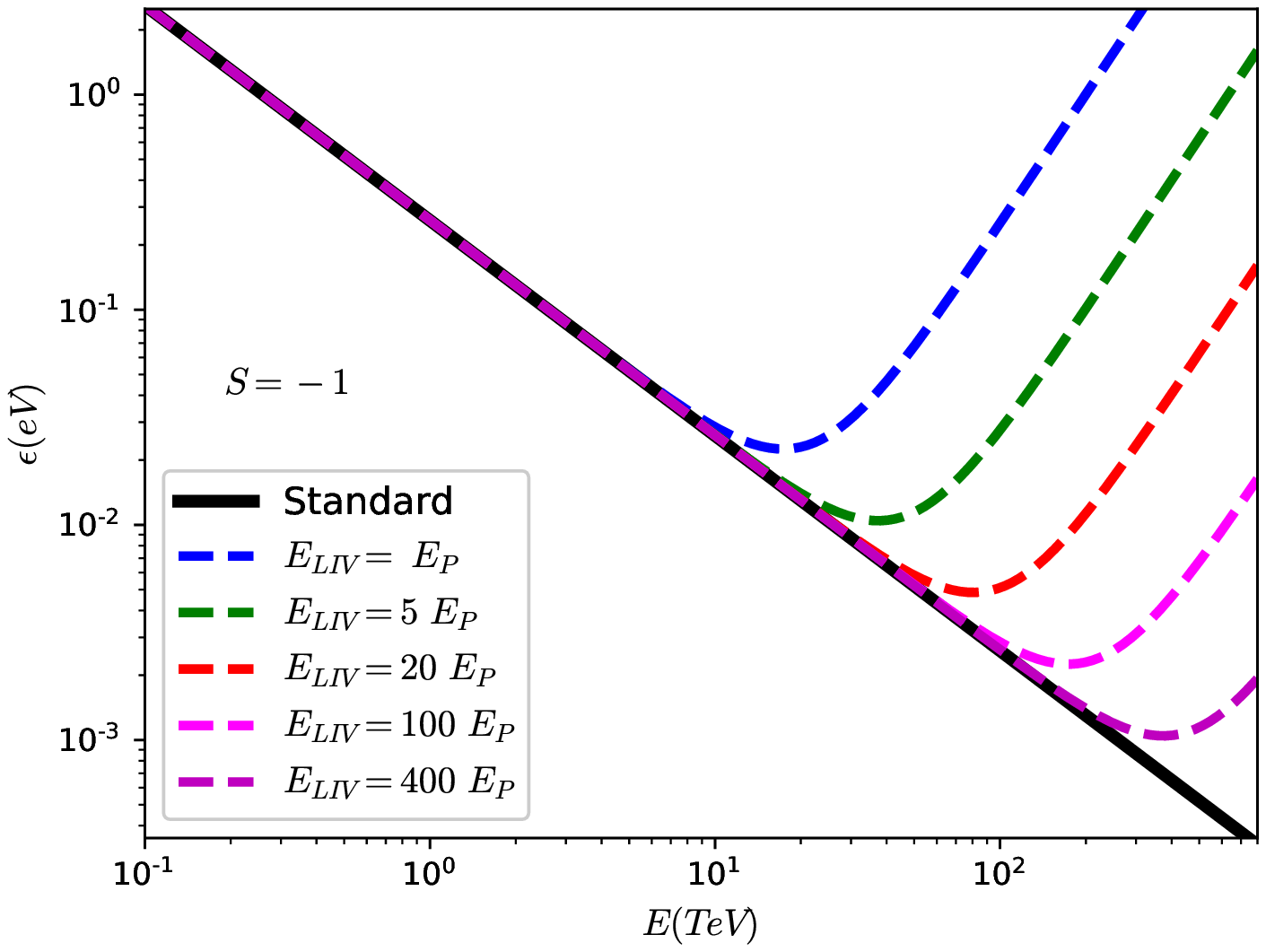}{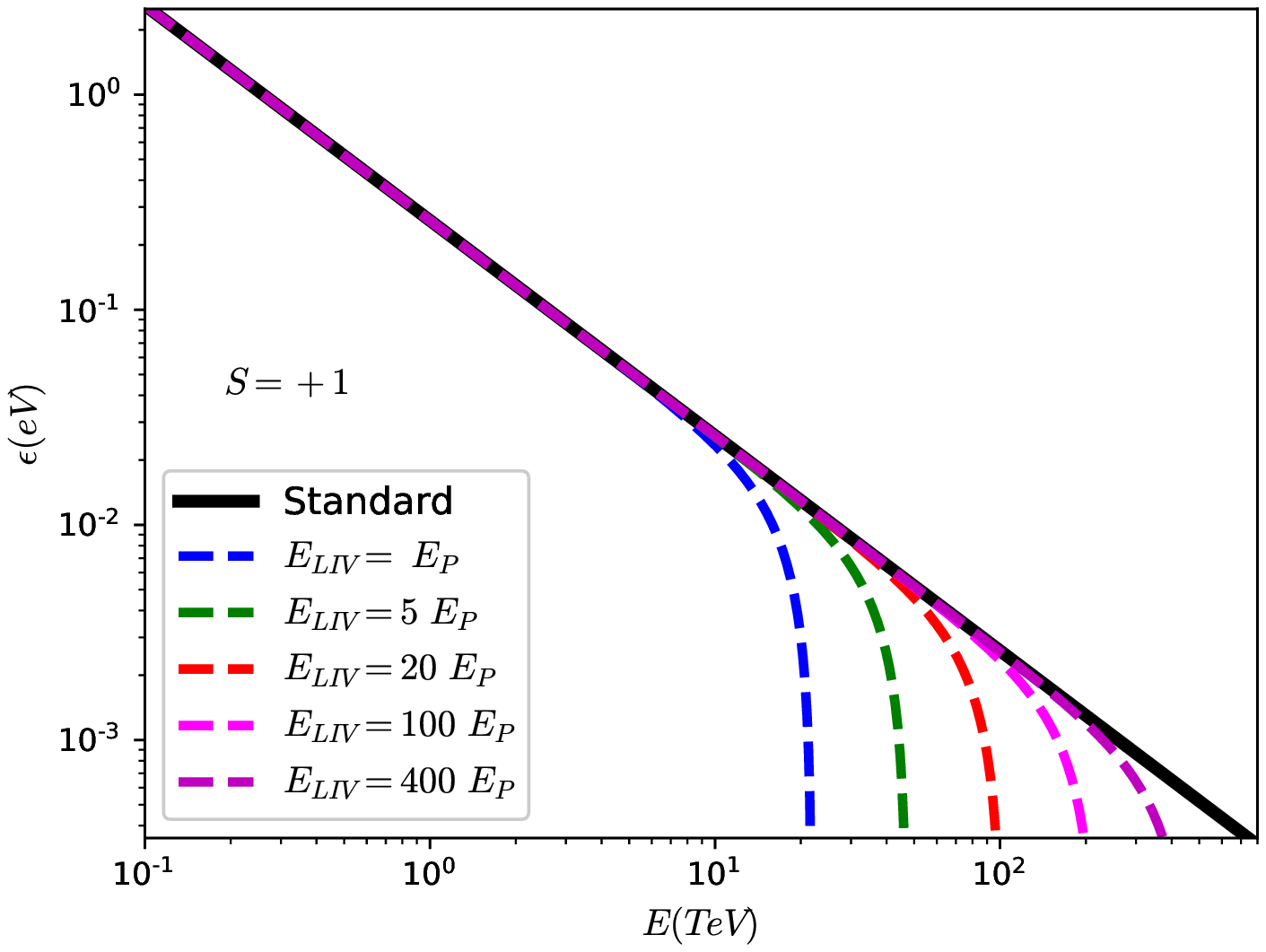}
\caption{Left panel: Photon target energy at threshold for pair-production as a function of $\gamma$-ray photon energy, 
for the subluminal case. The black solid line represents the case of standard QED and the dashed lines show the LIV-modified 
threshold for different values of $E_{LIV}$. Right panel: Same as the left panel, but for the superluminal case.
\label{fig:emin}}
\end{figure}

Based on the revised dispersion relation (\ref{dis}) with $n = 1$, the modified 
pair-production threshold energy $\epsilon_{min}$ can be written as \citep[e.g.,][]{Tavecchio16}:
\begin{equation}\label{th}
\epsilon_{min} = \frac{m^2 c^4}{E_{\gamma}} - ~  S ~ \frac{E_{\gamma}^2}{4 E_{LIV}}.
\end{equation}
Using equation (\ref{th}), the target photon energy threshold for pair-production as a function of the 
$\gamma$-ray photon energy for the subluminal and the superluminal cases is illustrated in Figure \ref{fig:emin}.

Also from equation (\ref{dis}), an effective mass term for photons can be defined as 
\citep[e.g.,][]{Liberati09,Tavecchio16}:
\begin{equation}\label{eff}
(m_{\gamma} \, c^2)^{2} \equiv  ~ S ~ \frac{ E^3 (1+z)^3}{ E_{LIV}}.
\end{equation}
Following \citep{Fairbairn14,Tavecchio16}, we assume that the functional form of the $\gamma-\gamma$ cross section
(as a function of the center-of-momentum energy squared $s$) remains unchanged by the LIV effect, and only the 
expression for $s$ is modified. The optical depth at the energy $E_{\gamma}$ and for $\gamma$-ray photons from 
a source at redshift $z_s$ can thus be evaluated as \citep{Fairbairn14,Tavecchio16}:
\begin{equation}\label{opa}
 \tau_{\gamma\gamma}(E_{\gamma}, z_s)  = \frac{c}{8 E_{\gamma}^2} \int_{0} ^{z_s} \frac{dz}{H(z) (1+z)^3} 
 \int_{\epsilon_{min}}^{\infty} \frac{n(\epsilon,z)}{\epsilon^2} \int_{s_{min}} ^{s(z) _{max} } [s - (m_{\gamma} \, c^2)^2] 
 \sigma_{\gamma\gamma}(s) ds,
\end{equation}
where $H(z) = H_{0} \sqrt{[\Omega_{m} (1+z)^3 + \Omega_{\Lambda}]}$,  $ s_{min} = 4 (m_{e} \, c^2)^2$ and $s(z) _{max}  = 
4 \epsilon E_{\gamma} (1+z) + (m_{\gamma} \, c^2)^{2}$. $n(\epsilon,z)$ is the EBL photon energy density as a function of 
redshift $z$ and energy $\epsilon$, and $\sigma_{\gamma\gamma}(s)$ is the total pair production cross-section as 
a function of the modified square of the center of mass energy $s = (m_{\gamma} \, c^2)^{2} + 2 \epsilon E_{\gamma} 
(1 - \cos(\theta))$, where $\theta$ is the angle between the soft EBL photon of energy $\epsilon$ and the VHE 
$\gamma$-ray photon. Obviously, when $E_{LIV} \longrightarrow \infty$, the standard relations are recovered.

By using equation (\ref{opa}) with the EBL model by \cite{Finke10}, we calculate the optical depth for VHE 
$\gamma$-ray photons from a source at redshift $0.6$. The comparison with the standard case (homogeneous EBL,
no LIV) and with the effect of EBL inhomogeneities (as discused in Section \ref{void}) is presented in 
Section \ref{EBLresults}. 
 
\section{Lorentz Invariance Violation: Compton scattering}
\label{livcom}

One of the most important fundamental high-energy radiation mechanisms is Compton scattering, the process by 
which photons gain or lose energy from collisions with electrons. In the Compton scattering processes, the energy 
of a scattered photon $E_{\gamma f}$ follows from momentum and energy conservation:
\begin{equation}\label{1c}
 \left(E_{\gamma i}/c, \overrightarrow{P} _{\gamma i}\right) + \left(E_{e i}/c, \overrightarrow{P} _{e i}\right) = 
 \left(E_{\gamma f}/c, \overrightarrow{P} _{\gamma f}\right) + \left(E_{e f}/c, \overrightarrow{P} _{e f}\right),
\end{equation}
which is assumed to still hold even in a Lorentz-invariance violating framework. 
In Equ. (\ref{1c}), $E_{\gamma i}$, $E_{\gamma f}$ and $E_{e i}$, $E_{e f}$  are initial and final energies for 
the photon and electron respectively, and $\overrightarrow{P}_{\gamma i}$, $\overrightarrow{P}_{\gamma f}$ and 
$\overrightarrow{P}_{e i}$, $\overrightarrow{P}_{e f}$  are initial and final momenta for the photon and electron,
respectively. To consider the LIV effect, we consider the first order correction $n = 1$ in the modified 
dispersion relation (\ref{dis}):
\begin{equation}\label{2}  
E_{\gamma}^{2} = p_{\gamma}^2 c^2  +  ~ S ~ \frac{E_{\gamma}^{3}}{E_{LIV}}, 
\end{equation}
As motivated in the introduction, and consistent with our treatment of LIV on the EBL opacity in Section \ref{livopa},
we consider LIV only for photons, not for electrons. Substituting for $E_{e f}$ using the standard electron dispersion
relation and momentum conservation (considering that in the electron rest frame, $p_{e, i} = 0$), the energy conservation 
part of Equ. (\ref{1c}) can be written as:
\begin{equation}\label{7} 
 E_{\gamma f} =  E_{\gamma i} + E_{e i} - \sqrt{  c^2 ( p_{\gamma i}  - p_{\gamma f})^2  + (m_{e}c^2)^2 }.
\end{equation}
Squaring and rearranging Equ. (\ref{7}), expressing all photon momenta in terms of energies using the dispersion
relation (\ref{2}) yields
\begin{eqnarray}\label{8x} 
 2 E_{\gamma i} E_{\gamma f} + 2 (E_{\gamma f} - E_{\gamma i}) m_e c^2 = S ~ 
 \left(\frac{E_{\gamma i}^3}{E_{LIV}}   +  \frac{E_{\gamma f}^3}{E_{LIV}} \right) + 2  ~ \mu ~  
 \sqrt{E_{\gamma i }^2 - S ~   \frac{E_{\gamma i}^3}{E_{LIV}}} \sqrt{E_{\gamma f }^2 
 - S ~  \frac{E_{\gamma f}^3}{E_{LIV}}}. 
\end{eqnarray}
where $\mu = \cos \theta$ is the cosine of the scattering angle in the electron rest frame. 
In the limit $E_{LIV} \gg E_{\gamma}$, the square-root expressions in Equ. (\ref{8x}) can be simplified to
\begin{equation}
\sqrt{E_{\gamma}^2 - S ~  \frac{E_{\gamma}^3}{E_{LIV}} } \approx E_{\gamma} \left( 1 - S ~ \frac{E_{\gamma}}{2 E_{LIV}} \right).
\label{squareroots}
\end{equation}
Thus, to lowest order in $ E_{\gamma} / E_{LIV}$, Equ. (\ref{8x}) can be written as: 
\begin{eqnarray}\label{9x} 
 2 E_{\gamma i} E_{\gamma f} + 2 (E_{\gamma f} - E_{\gamma i}) m_e c^2 = S ~  \left(\frac{E_{\gamma i}^3}{E_{LIV}}  
 +  \frac{E_{\gamma f}^3}{E_{LIV}} \right) 
 + 2  \mu E_{\gamma i } E_{\gamma f } \left( { 1 - S ~ \frac{E_{\gamma i}}{2 E_{LIV}}} - 
 S ~ \frac{E_{\gamma f}}{2 E_{LIV}} \right). 
\end{eqnarray}
Equ. (\ref{9x}) is solved numerically to find the scattered photon energy $ E_{\gamma f}$ as a function of
initial photon energy $E_{\gamma i}$ and scattering angle $\theta = \cos^{-1} \mu$. Results are presented 
in Section \ref{KNresults}. \\
From QED, the Klein-Nishina cross-section $\sigma_{KN}$ can be written as:
\begin{equation}\label{KN} 
\sigma_{KN} = \int \frac{d\sigma_{KN}}{d\Omega} d\Omega = \int \frac{r_e^2}{2} \left( \frac{E_{\gamma f}}{E_{\gamma i}}
\right)^2 \> \left(\frac{E_{\gamma i}}{E_{\gamma f}}  +  \frac{E_{\gamma f}}{E_{\gamma i}} - \sin^2 \theta \right) d\Omega,
\end{equation}
where  $\frac{d\sigma_{KN}}{d\Omega}$ is the differential Klein-Nishina cross section and $d\Omega$ is the solid angle,
and $r_e$ is the classical electron radius.

As for our considerations of the LIV effect on the $\gamma-\gamma$ opacity, we assume that the functional dependence of the
Klein-Nishina cross section on the incoming and scattered photon energies remains unaffected. Thus, in order to modify the 
Klein-Nishina cross-section considering the LIV effect, we use the scattered photon $E_{\gamma f}$ from the solution of 
Equ. (\ref{9x}) in the Klein-Nishina formula (\ref{KN}) and integrate numerically. The results of this integration compared 
with the standard QED case are presented in Section \ref{KNresults}.
\section{Results and Discussion}
\label{Results}
In this section, we present the results for representative test cases for the LIV effect on the cosmic $\gamma-\gamma$
opacity, compared standard Lorentz-invariance case and the suppression of the opacity due to EBL inhomogeneities, and 
on the Compton scattering process, compared to the standard-model case. 
\subsection{\label{EBLresults}EBL Absorption}
\begin{figure}[ht!]
\plottwo{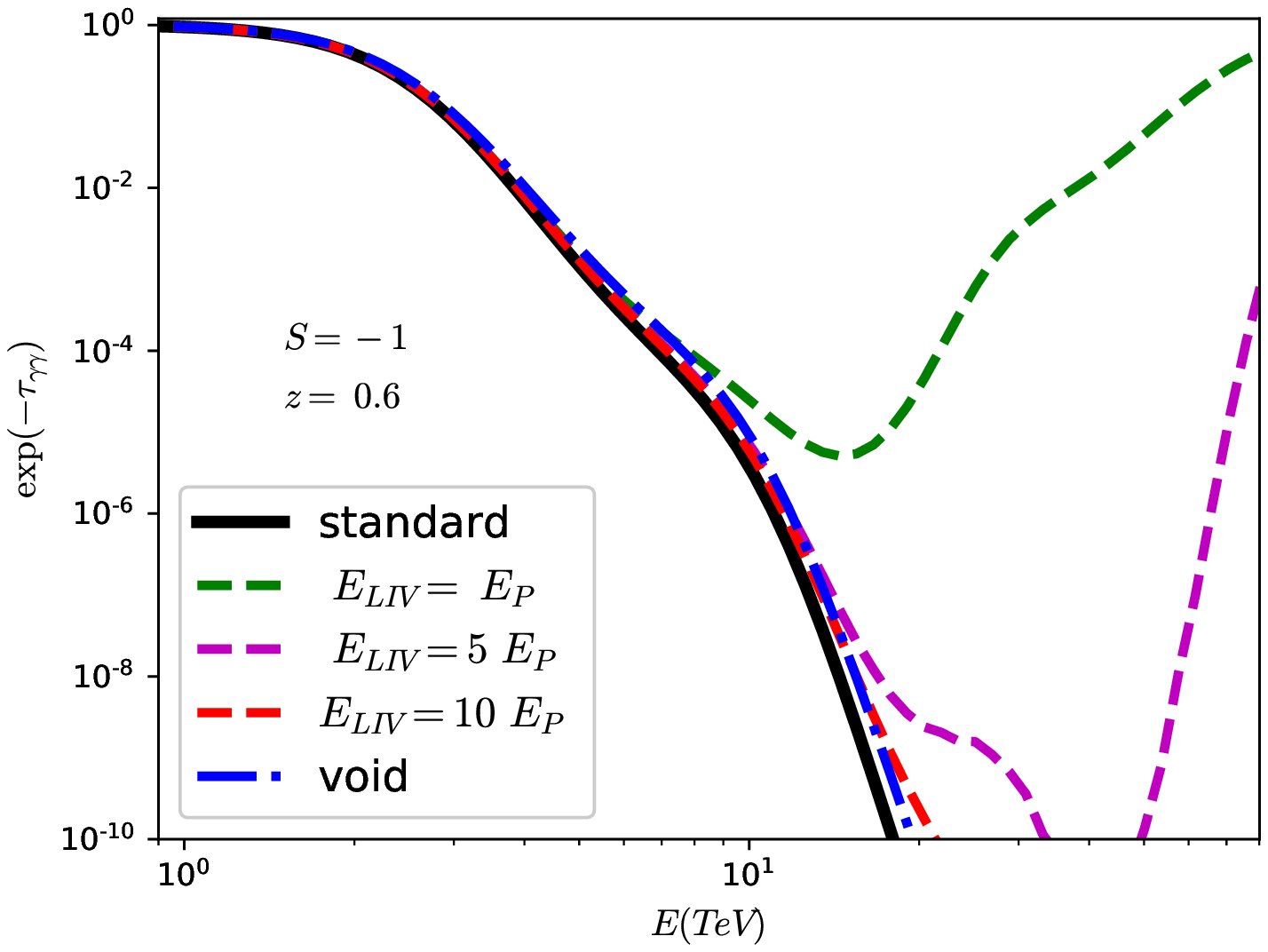}{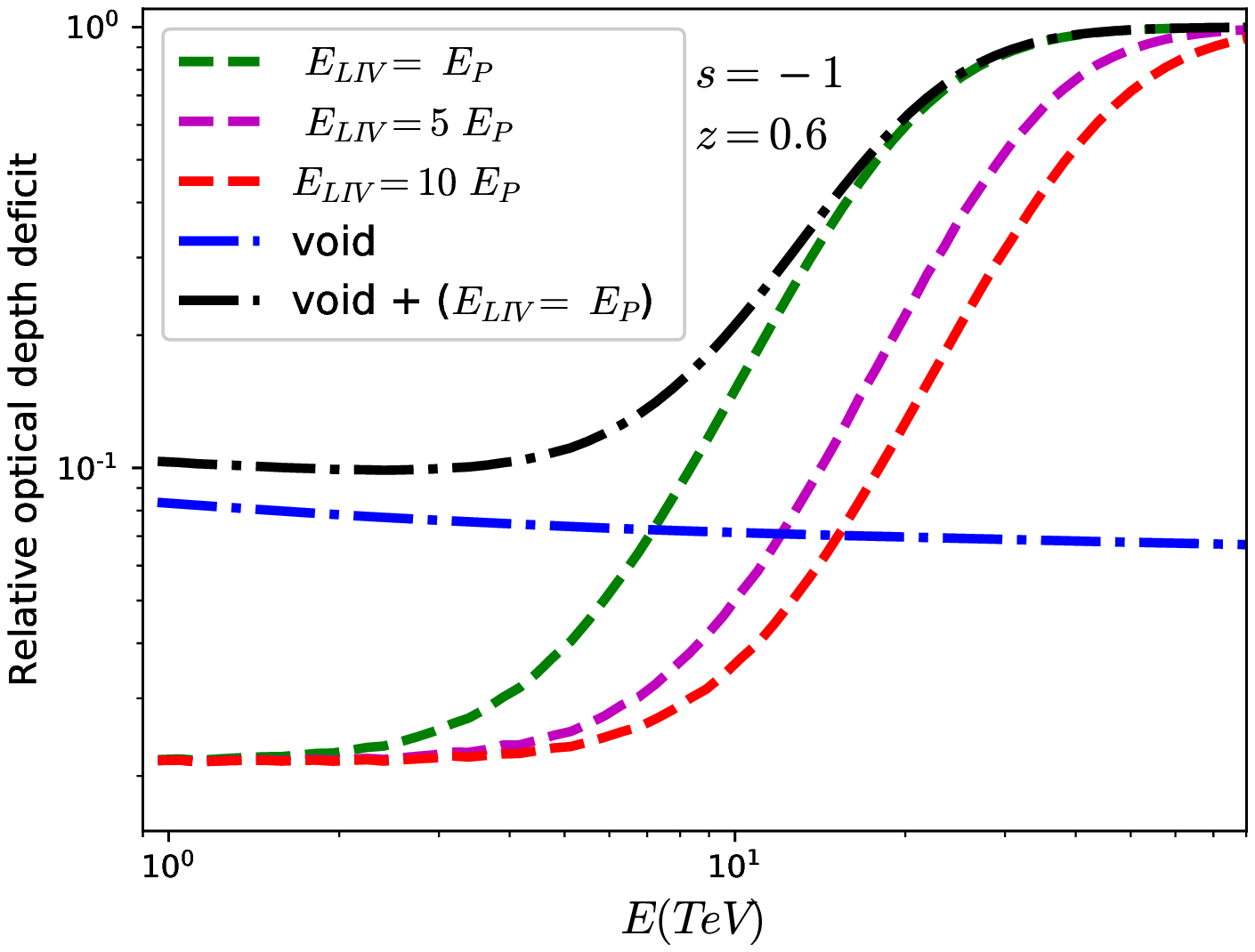}
\plottwo{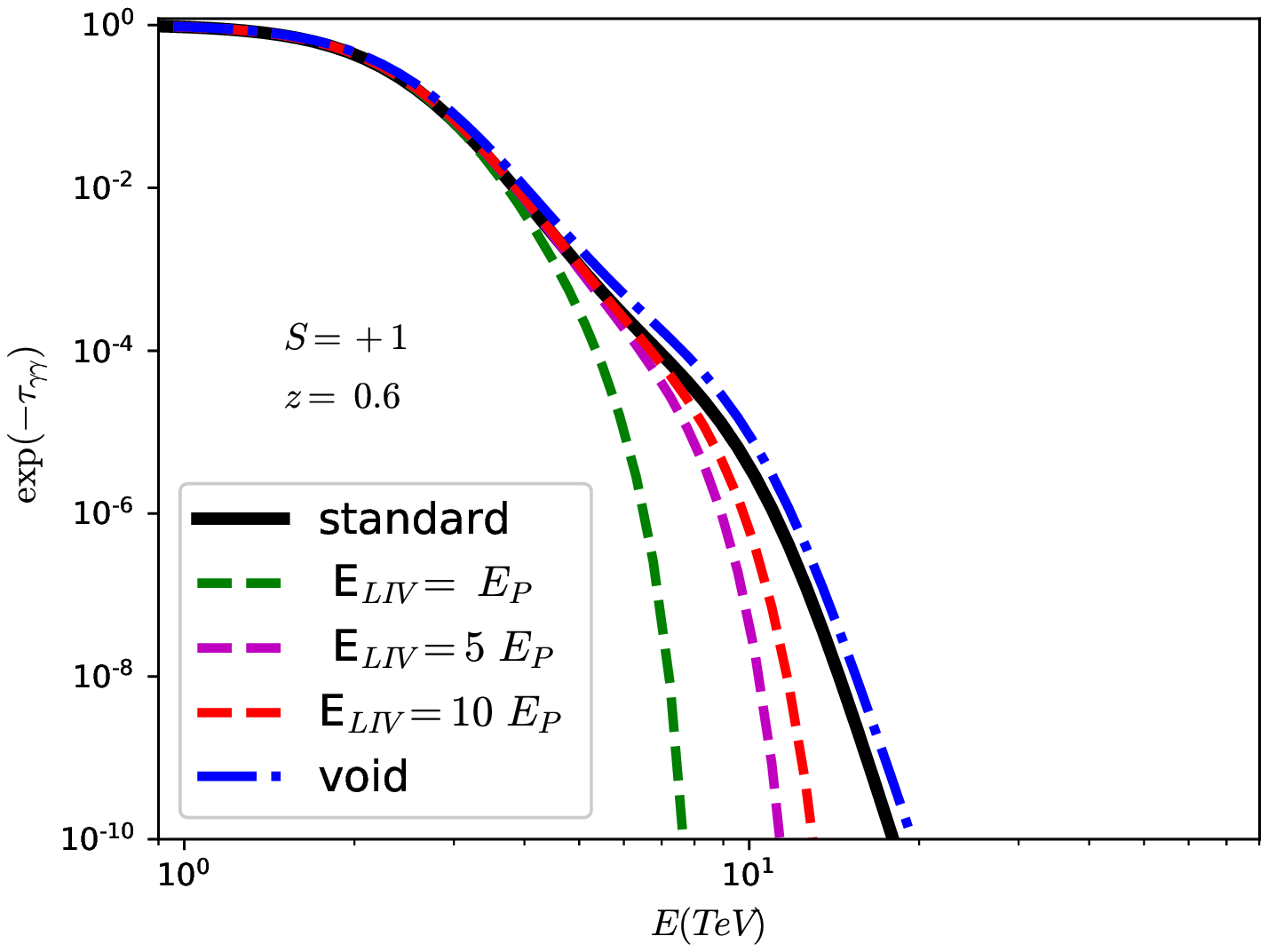}{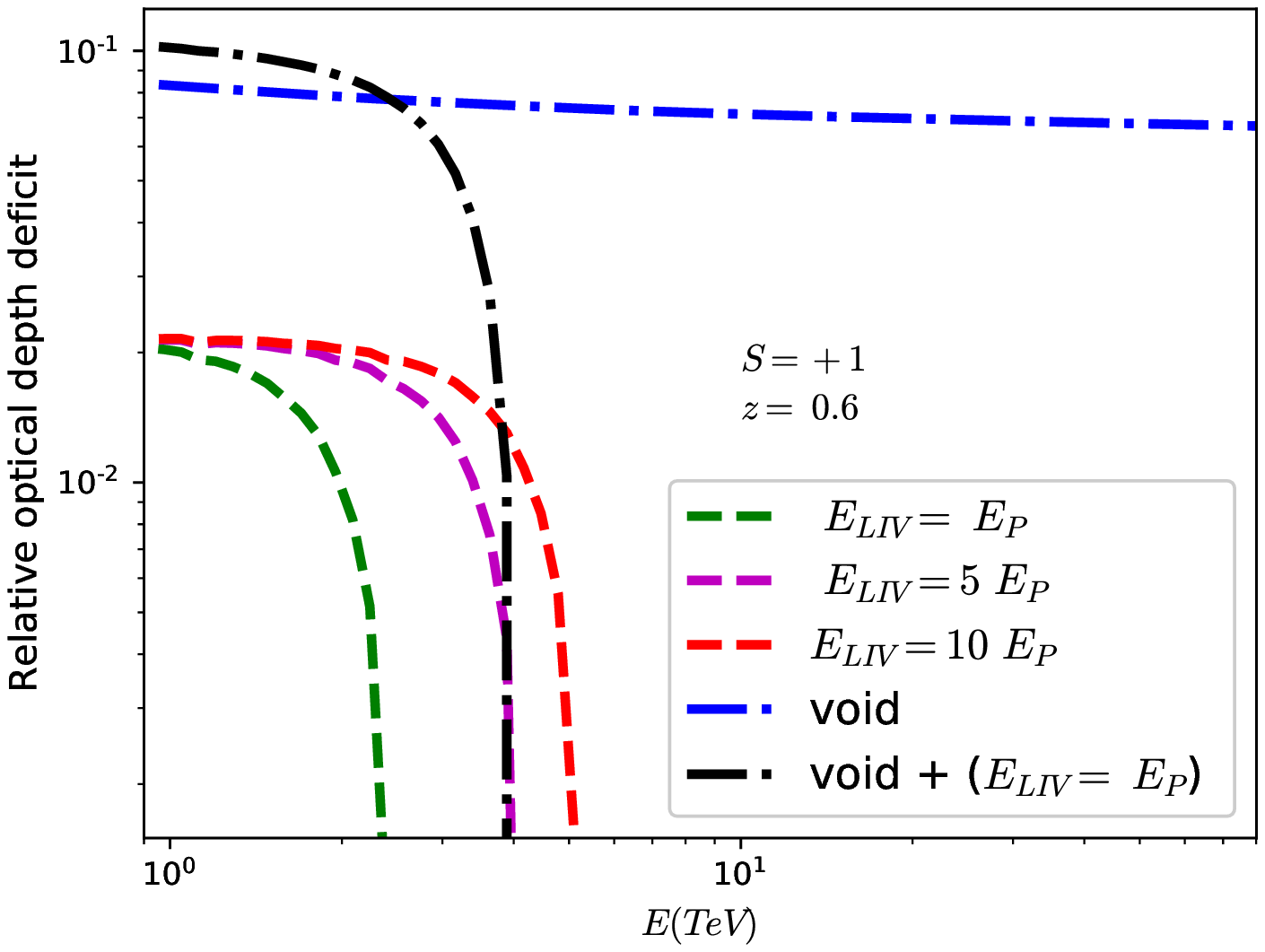}
\caption{Left panels:  Absorption coefficient $\exp(- \tau_{\gamma\gamma})$ as a function of energy for VHE $\gamma$-rays 
from a source at redshift $z_s = 0.6$, using the EBL model of \cite{Finke10}. The black solid line represents the case 
of standard QED; the dashed lines show the LIV-modified coefficient for different values of $E_{LIV}$, for the subluminal 
case (top panel) and the superluminal case (bottom panel). 
The blue dot-dashed line represents the case of standard QED and EBL energy density calculated by considering an 
accumulation of 10 voids of typical sizes with radius $R = 100 \, h^{-1}$~Mpc along the line of sight, centered at 
redshift $z_v= 0.3$.
Right panels: Relative optical depth deficit as a function of energy for VHE $\gamma$-rays for 
the same cases as in the left panel. The Relative optical depth deficit is defined as 
($ 1 - {\tau_{\gamma\gamma}^{DFS}} /{\tau_{\gamma\gamma} ^{Stand.}} $), where $\tau_{\gamma\gamma}^{Stand. }$ represents 
the optical depth calculated in standard QED and using the homogeneous EBL energy density distribution, and 
$\tau_{\gamma\gamma}^{DFS}$ represents the optical depth calculated including the effects of cosmic voids (blue dashed-dot line) 
or of LIV (dashed lines). The black dot-dashed line represents the relativie optical depth deficit due to the 
combined effect of LIV and EBL inhomogeneities.
\label{fig:opaliv}}
\end{figure}

To study the opacity or transparency of the Universe to VHE $\gamma$-ray photons from distant sources (e.g. blazars) due to 
their interaction with intergalactic EBL photons, we compare the effects of the EBL inhomogeneities due to the presence 
of cosmic voids to those of the LIV effect. Figure (\ref{fig:opaliv}) shows the absorption coefficient 
$\exp(- \tau_{\gamma\gamma})$ as a function of energy for VHE-gamma rays from a source at redshift $z_s = 0.6$. 
The standard-model QED case is represented by the black solid line. The impact of an EBL underdensity (for parameters
as used in Fig. \ref{fig:EBL}) is illustrated by dot-dashed lines and the LIV effect is represented by dashed lines 
for different values of the chracteristic LIV energy scale $E_{LIV} = E_P/\xi_1$. Note that the standard case without 
LIV is recovered for $E_{LIV} \longmapsto \infty$.
\vspace{-5.0mm} 

The reduction of the EBL $\gamma-\gamma$ opacity due to plausible EBL inhomogeneities is only of the order of $\lesssim
10$~\% and decreases with energy. The LIV effect is negligibly small for energies below about 5 TeV, but the cosmic opacity 
for VHE $\gamma$-rays with energies $\gtrsim 10$~TeV can be strongly reduced for the subluminal case and increased for the 
superluminal case. Therefore, if LIV is described by the subluminal dispersion relation ($S = -1$), one may expect VHE 
$\gamma$-ray photons beyond 10 TeV to be observable even from distant astrophysical sources.\\
However, the spectral hardening of several observed VHE gamma-ray sources with energy from 100 GeV up to a few TeV 
(e.g. PKS 1424+240) still remains puzzling. Compared to the \cite{Finke10} EBL absorption model for an object at a 
redshift of $z_s \sim 0.6$, the opacity would have to be reduced by $\gtrsim 60$~\% in order to explain the spectral hardening
of the VHE spectrum of PKS 1424+240 with standard emission mechanisms. Even if we consider the combined effects 
of EBL underdensities and LIV, as represented by the solid line in the right panel of Figure (\ref{fig:opaliv}), the 
relative optical depth $\tau{\gamma\gamma}$ deficit is only around 10~\% in the energy range from hundred of GeV to 
a few TeV.
\subsection{\label{KNresults}Compton scattering}
\begin{figure}[ht!]
\plottwo{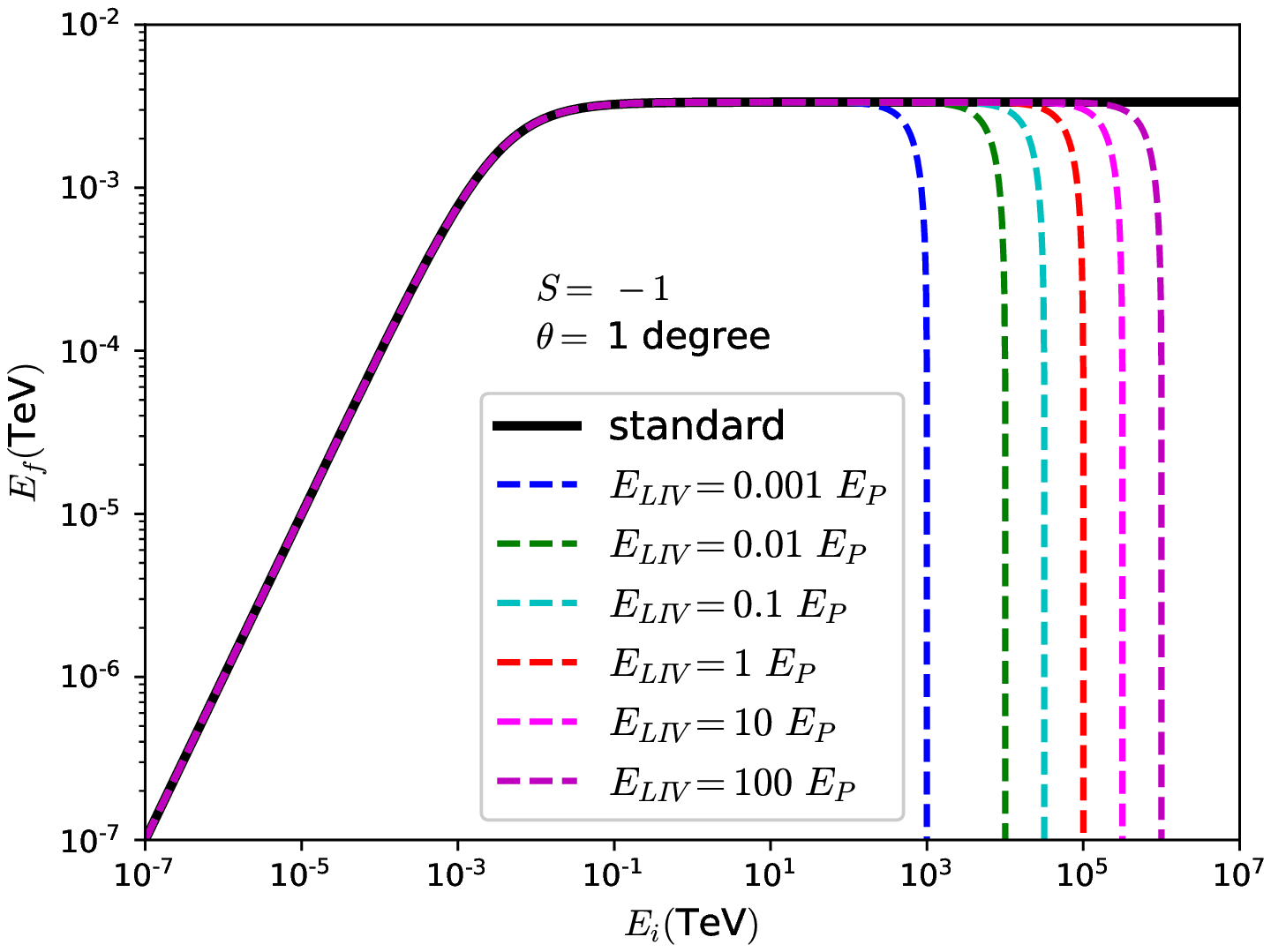}{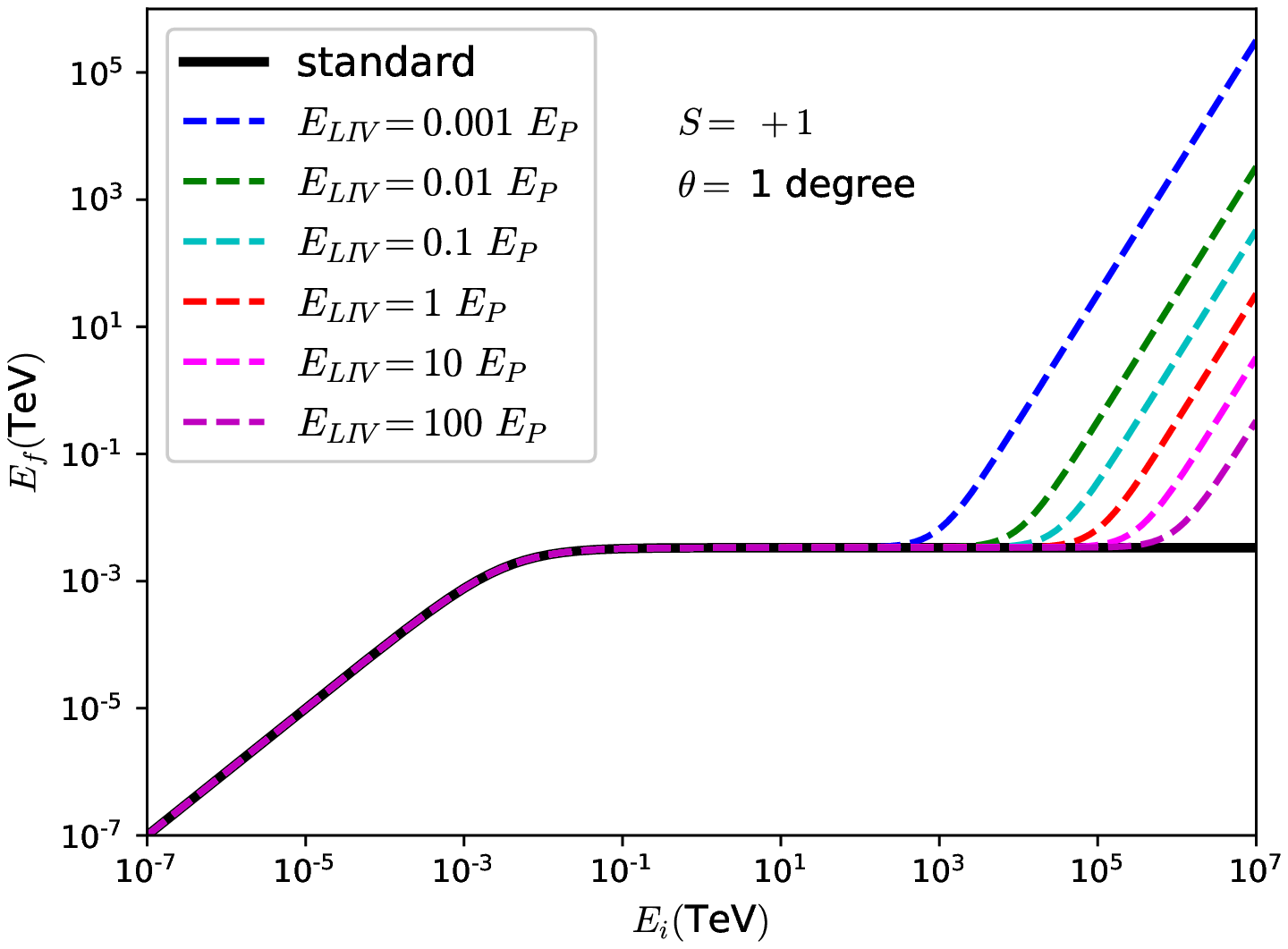}
\plottwo{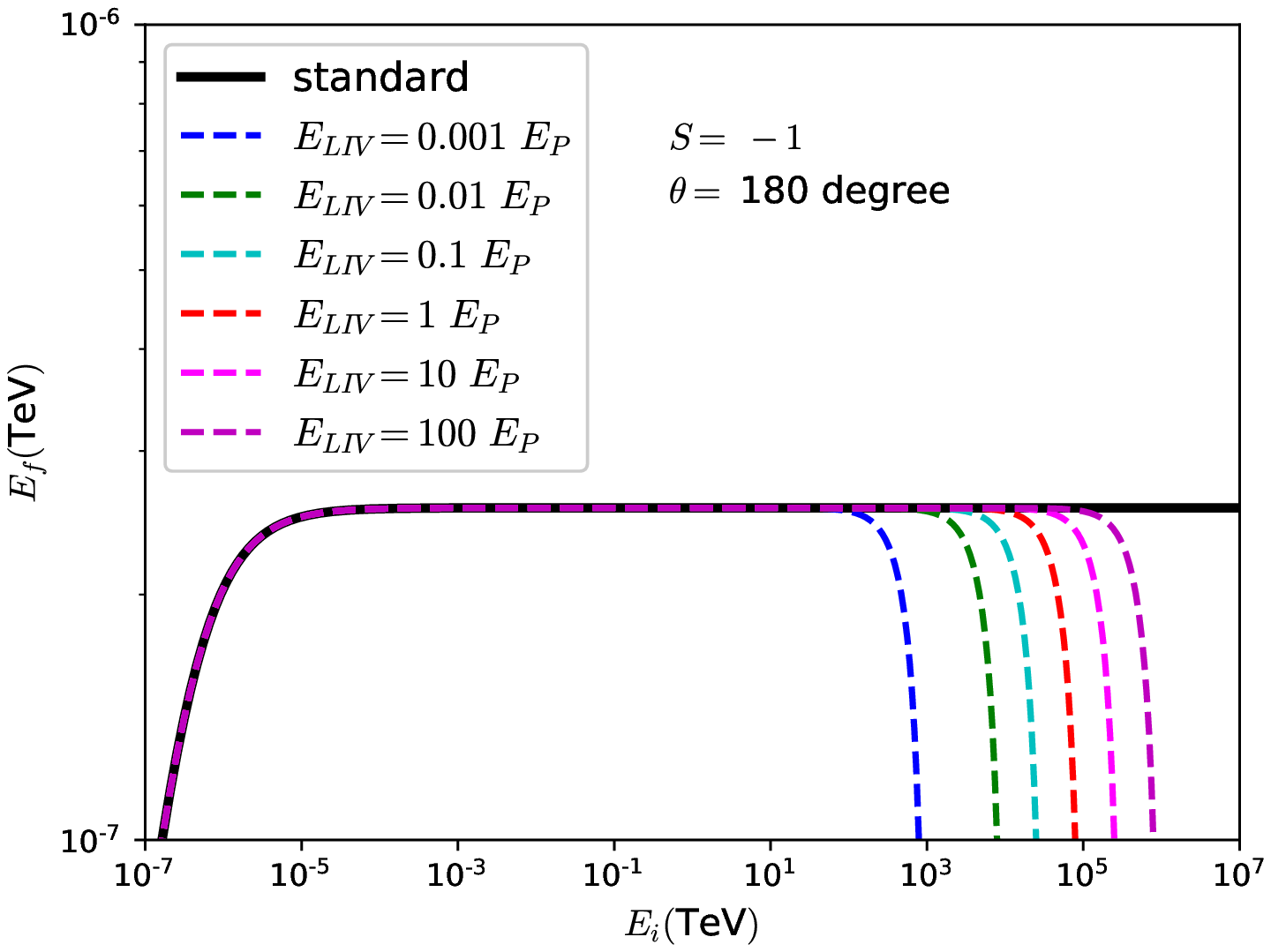}{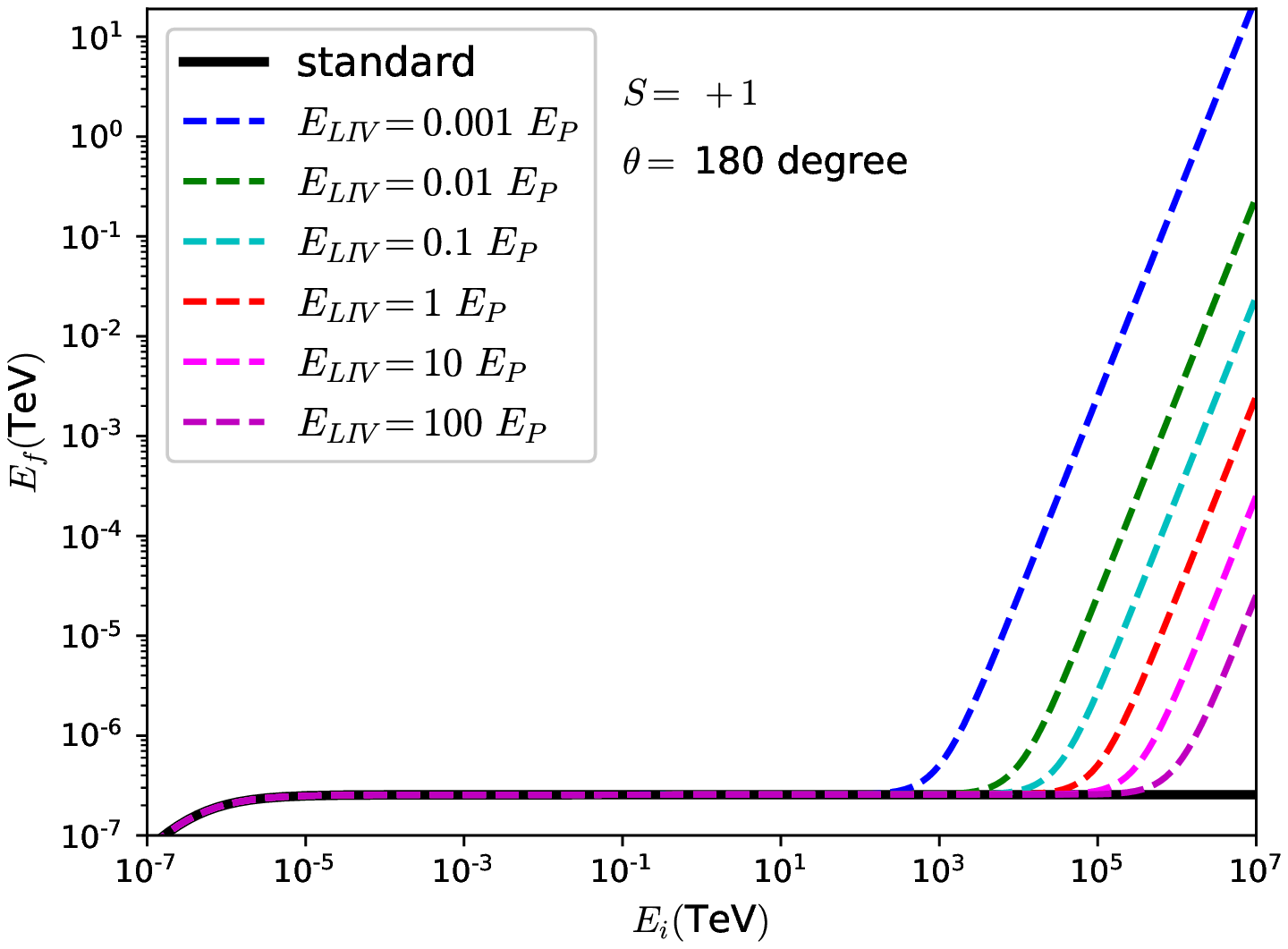}
\plottwo{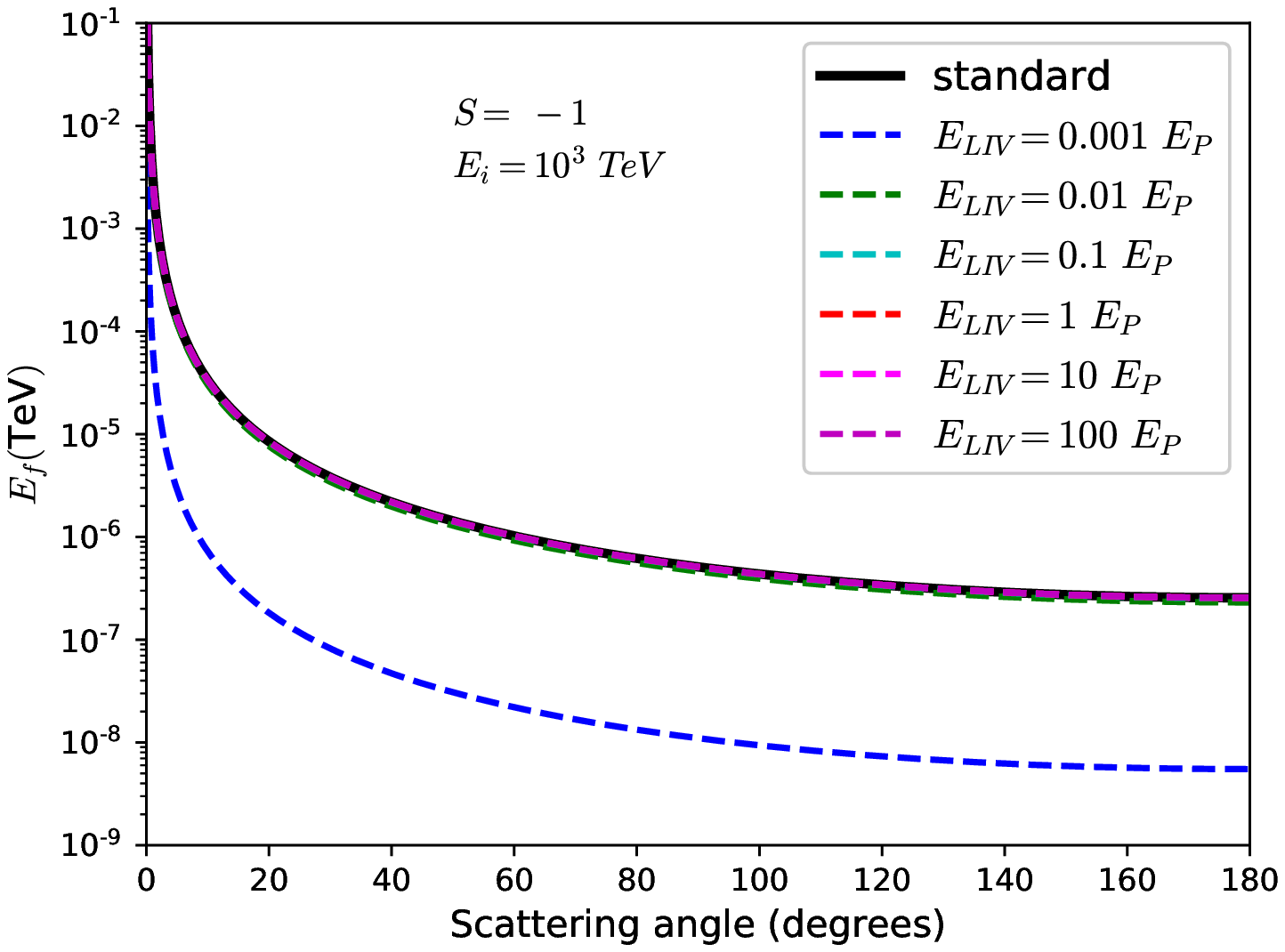}{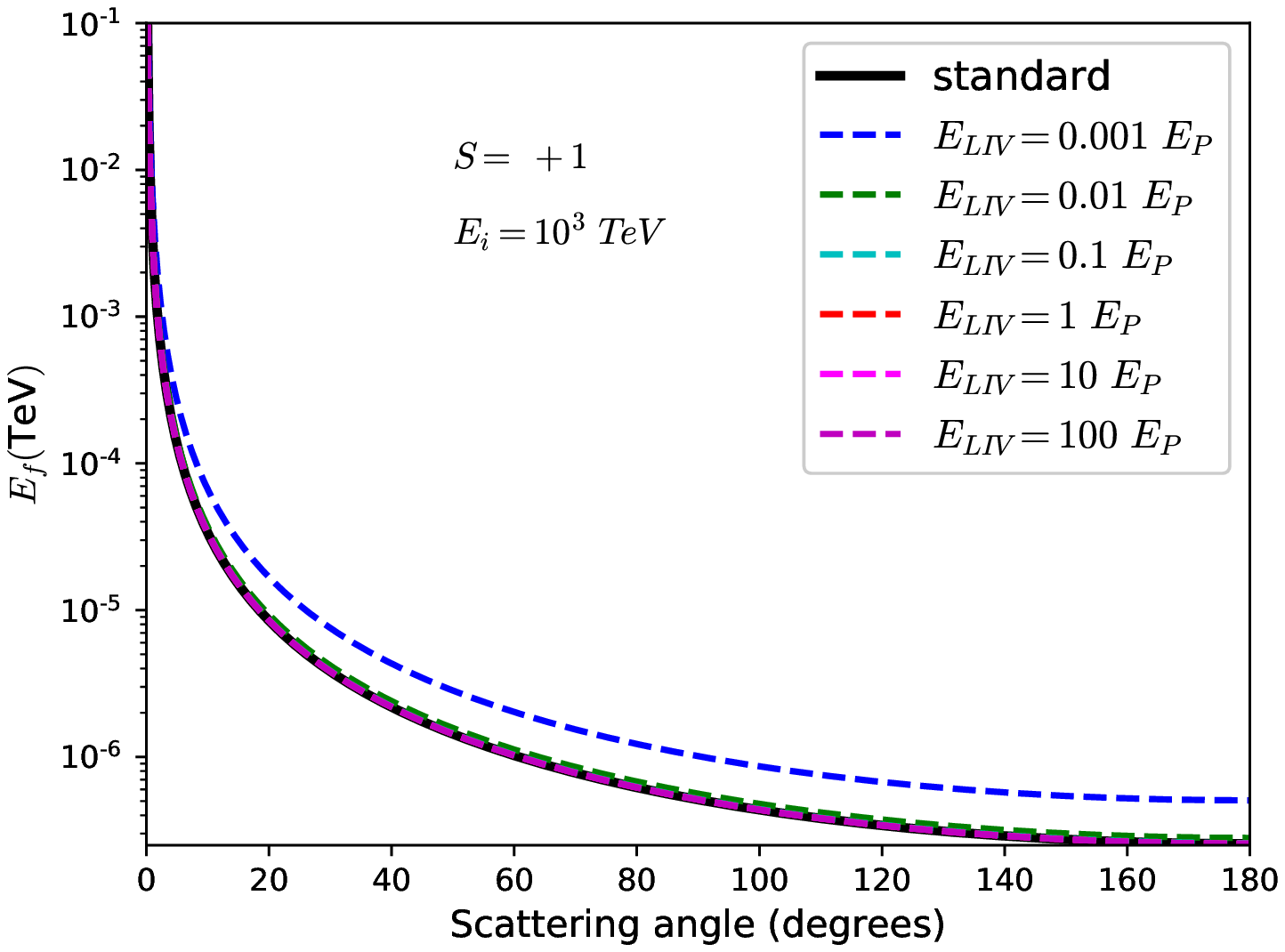}
\caption{Top and middle panels: Scattered photon energies $E_{\gamma,f}$ as a function of incoming photon energy 
$E_{\gamma,i}$, for scattering angles of $1$ and $180$ degrees, respectively. 
The black solid line represents the case of standard QED; the dashed lines show the LIV effect for different 
values of $E_{LIV}$, for a subluminal case (left) and superluminal case (right).
Bottom panels: Scattered photons energies $E_{f}$ vs. scattering angle, for an incoming photon energy 
of $E_{i} = 1$~PeV in the subluminal case (left) and superluminal case (right). The black solid line represents the case 
QED; the dashed lines illustrate the LIV effect for different values of $E_{LIV}$.
\label{fig:EFEI}}
\end{figure}
The LIV effect on the Compton scattering process has been evaluated as described in Section \ref{livcom}. 
To assess the importance of LIV signatures, we have evaluated this effect for a large range of values of $E_{LIV}$.
All calculations are done in the electron rest frame. 

Figure \ref{fig:EFEI} illustrates the effect of LIV on the scattered photon energies as a function of the 
incoming photon energies $E_{i}$ for two representative scattering angles ($1$ and $180$ degrees --- top and middle panels) 
for different values of $E_{LIV}$, as well as the scattered photon energies as a function of the scattering angle $\theta$ 
for one representative incoming photon energy ($10^3$~TeV --- bottom panels). The subluminal cases are illustrated in the 
left, the superluminal cases in the right panels. In the standard QED case (black solid curves), the kinematic constraints 
(recoil) lead to the well-known levelling-off of the scattered photon energies at a value of $E_{\gamma,f} \sim m_e c^2 / 
(1 - \cos\theta)$.
\begin{figure}[ht]
\plottwo{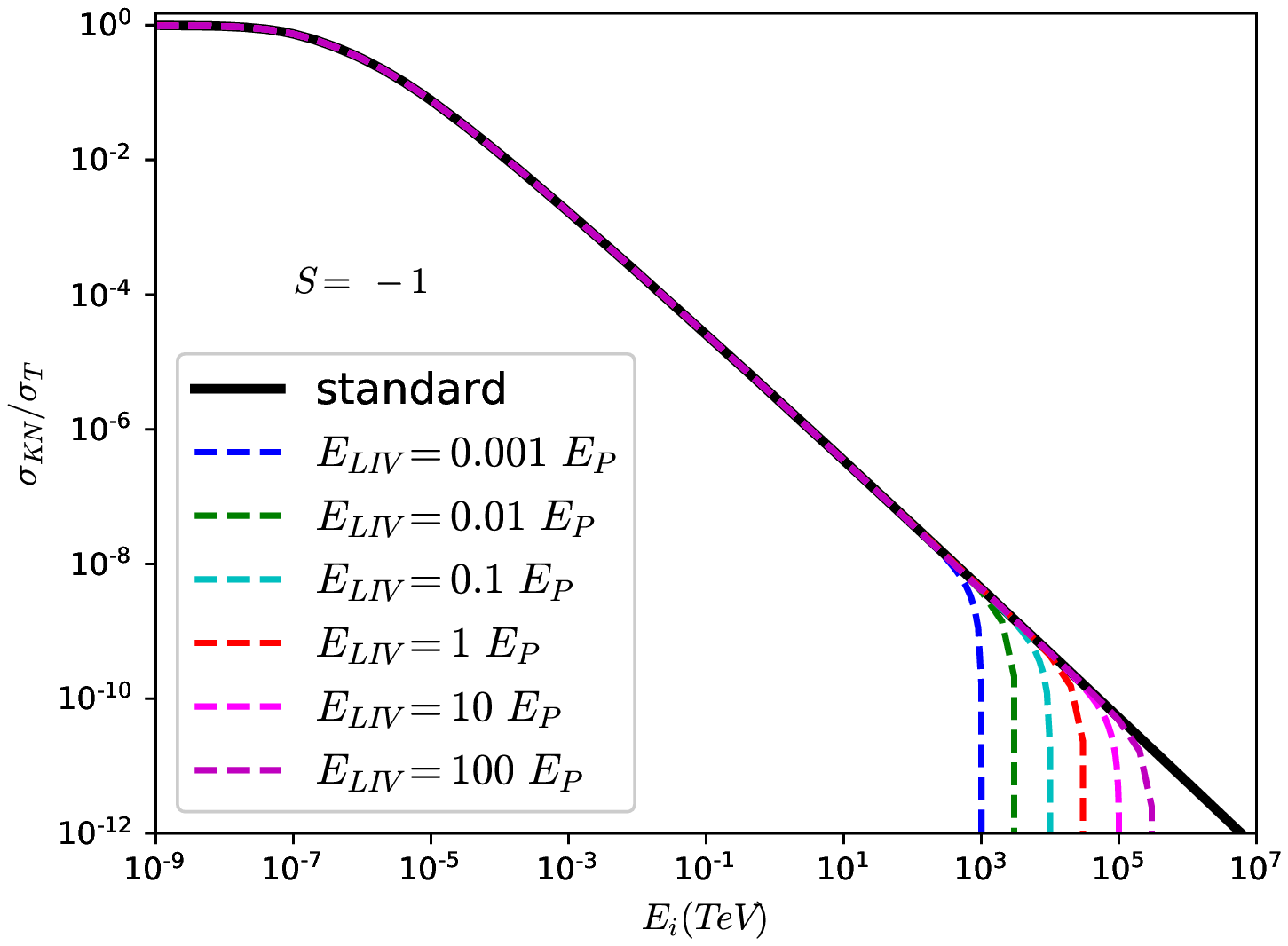}{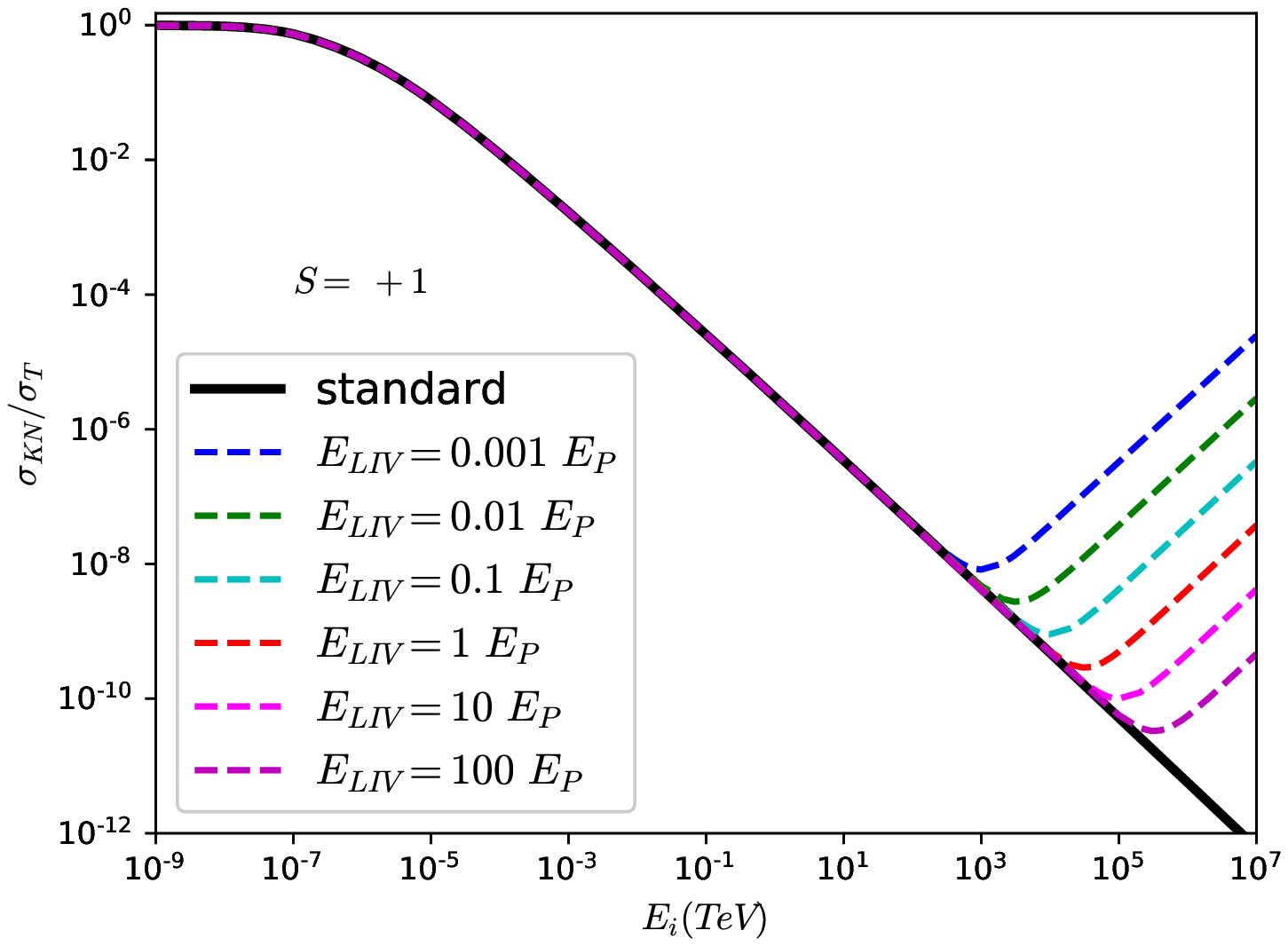}
\caption{Total Klein-Nishina cross-section $\sigma_{KN}$ (in the units of $\sigma_{T}$) as a function of 
the incoming photon energy $E_{\gamma,i}$. The black solid line represents the case of QED; the dashed lines 
show the LIV-modified Klein-Nishina cross-section for different values of $E_{LIV}$, for the subluminal
(left panel) and superluminal case (right panel). 
\label{fig:KN}
}
\end{figure}
In Figure \ref{fig:KN} we illustrate the LIV effect on the total Klein-Nishina cross-section $\sigma_{KN}$ 
(in units of $\sigma_{T}$), plotted as a function of the incoming photon energy $E_{\gamma,i}$. The black 
solid line represents the case of standard QED and the dashed lines show the modified Klein-Nishina cross-section 
for different values of $E_{LIV}$, calculated as described in Section \ref{livcom}. Again, the subluminal and 
 superluminal cases are illustrated in the left and right panel, respectively. 

From Figures (\ref{fig:EFEI}) and (\ref{fig:KN}) we see that LIV signatures in the Compton scattering processes 
are expected to be important only for very large incoming photon energies, $E_{\gamma,i} \gtrsim 1$~PeV. In the 
superluminal case, the scattered photon energies are larger than expected in the standard case, while in the subluminal
case, the scattered photon energies are further reduced. Although the impact of this effect on the scattered photon 
energy is large for photons with energy $E_{\gamma,i}  > 10$~PeV, even in the superluminal case the scattered photon 
energy $E_{\gamma,f}$ is still much smaller than the incoming photon energy $E_{\gamma,i}$. This indicates 
that the electron recoil effect is still substantial, as expected, but strongly reduced/increased compared to 
standard-model kinematics, in the superluminal/subluminal case, respectively. 
Equally, at energies $E_{\gamma,i} \gtrsim 1$~PeV the Klein-Nishina cross section gradually recovers from the 
standard-model Klein-Nishina suppression (which sets in at $E_{\gamma,i} \sim m_e c^2$) in the superluminal case, 
but is expected to remain suppressed to $\sigma_{KN} \lesssim 10^{-6} \, \sigma_T$ for photon energies below 
$\sim 1$~EeV (in the electron rest frame) for any plausible choice of $E_{LIV}$. In the subluminal case, 
Compton scattering of photons at energies $E_{\gamma, i} \gtrsim 1$~PeV is expected to be strongly suppressed,
far beyond the standard-QED Klein-Nishina suppression. 
\\

\section{Summary and Conclusions}
\label{Summary}
We have presented calculations of the modification of the EBL $\gamma-\gamma$ opacity for VHE $\gamma$-ray photons from 
sources at cosmological distances, by considering two effects: the impact of under-densities (voids) along the line of 
sight to the source and the LIV effect. For the LIV effect, we considered both a subluminal and a superluminal
modification of the dispersion relation for photons. We found that the reduction of the optical depth due to the 
existence of cosmic voids is insignificant for realistic parameters of the void and is thus insufficient to explain 
the unexpected spectral hardening of the VHE spectra of several blazars. The effect of LIV becomes important only at 
$\gamma$-ray energies above $\sim 10$~TeV, where the $\gamma\gamma$ interaction threshold is increased and consequently,
the EBL opacity is reduced in the subluminal case. The opposite effect (reduced pair production threshold and increased
EBL opacity) results in the superluminal case. The effect is negligible for VHE spectra 
in the range $\sim 100$~GeV -- a few TeV. However, these results suggest that, if LIV is manifested by a subluminal
modification by the photon dispersion relation, VHE $\gamma$-ray sources may be detectable at cosmological redshifts 
$z \gtrsim 1$ at energies $E \gtrsim 10$~TeV, as the EBL opacity at those energies may be greatly reduced compared 
to standard-model predictions. Observations with the small-size telescopes of the future Cherenkov Telescope Array 
\citep[CTA:][]{Acharya13} --- and its predecessors, such as the ASTRI \citep[Astrofisica con Specchi a Technologia 
Replicante Italiana:][]{Vercellone16} array --- will provide excellent opportunities to test this hypothesis. 

We have presented, to the authors' knowledge for the first time, detailed calculations of the effect of LIV on the 
Compton scattering process. As for $\gamma\gamma$ absorption, we considered both subluminal and superluminal 
modifications to the photon dispersion relation. In the superluminal case, we find that for incoming photon 
energies of $E_{\gamma,i} \gtrsim 1$~PeV in the electron rest frame, both the electron recoil effect and the 
Klein-Nishina suppression of the scattering cross section are reduced compared to standard-model expectations. 
This may suggest that Compton scattering at ultra-high energies may overcome the suppression due to the standard 
Klein-Nishina effect and possibly lead to the production of $>> 1$~PeV photons through inverse Compton scattering. 
However, it is unlikely that this effect is of relevance to realistic astrophysical environments. Such scattering 
would require electrons of energies $E_e \gg 1$~PeV. In spite of the recovery at ultra-high energies, the Compton 
cross section is still suppressed by several orders of magnitude compared to the Thomson cross section. Hence, for 
any realistic magnetic field value in an astrophysical source, if electrons are actually accelerated to $E_e \gg 1$~PeV,
or produced as secondaries in ultra-high-energy muon decay processes, they are likely to lose their energy radiatively 
via synchrotron radiation rather than Compton scattering. In the subluminal case, both the reduction of scattered 
photon energies and the Klein-Nishina cut-off of the cross section are further enhanced by the LIV effect, rendering
Compton scattering at ultra-high energies even less efficient than due to the standard Klein-Nishina effects.

\section{Acknowledgments}

We thank the anonymous referee for a quick review and helpful suggestions. 
The work of M.B. is supported through the South African Research Chair Initiative of the National 
Research Foundation\footnote{Any opinion, finding and conclusion or recommendation expressed in 
this material is that of the authors and the NRF does not accept any liability in this regard.} 
and the Department of Science and Technology of South Africa, under SARChI Chair grant No. 64789. 



\end{document}